# Reproducible Monolayer MoS$_2$ Devices Free of Resist Contamination by Gold Mask Lithography

Yumeng Liu[1], Yizhuo Wang[1], Zhengfang Fan[1], Jianyong Wei[1], Shuwen Guo[1], Zhijuan Su[2*], Yaping Dan[1*]

[1] University of Michigan – Shanghai Jiao Tong University Joint Institute, Shanghai Jiao Tong University, Shanghai, 20240 China

[2] Global Institute of Future Technology, Shanghai Jiao Tong University, Shanghai, 20240 China

Email: yaping.dan@sjtu.edu.cn, zhijuan.su@sjtu.edu.cn

Abstract

Atomically thin MoS$_2$ is a promising material for field-effect transistors (FETs) and electronic devices. However, traditional photolithographic processes introduce surface contamination to 2D materials, leading to poor electrical contacts when metals are deposited. In this work, we present a novel fabrication method using gold as a mask for patterning and etching, which protects 2D materials from contamination in the metal contact region. This technique enabled the fabrication of monolayer MoS$_2$ transistors with clean gold contacts. Additionally, we achieved MoS$_2$ devices with Ohmic contacts, mass-produced traditional lithography and gold mask lithography devices (100 of each), with the latter having a much higher current statistical variance than the former, which demonstrated the effectiveness of this method for contamination-free 2D transistors and potential applications in integrated circuits

Two-dimensional (2D) semiconducting materials, such as transition metal dichalcogenides (TMDs), have emerged as a promising semiconductor to potentially replace silicon for next-generation integrated circuits[1], [2], [3], [4], [5], [6]. Photolithography is an indispensable process for integrated circuits in which polymeric photoresists are used. After development, residues from these polymeric photoresists are often left on the underlying Si[7], [8]. Excessive harsh processes such as oxygen plasma are employed to remove the residues before metal is deposited to make good Ohmic contacts with Si[9], [10]. For atomically thin 2D semiconductors, it is well known that organic photoresist will form a thin layer (~ 1nm) of residue that is chemically attached to the 2D monolayers and cannot be removed by widely used solvents[9], [11], [12]. A harsh process like oxygen plasma often damages the 2D semiconductors and poor metal-semiconductor contacts will form[13], [14], [15], [16].

To address this challenge, in this work we propose a novel fabrication method that employs gold as a protective mask for atomically thin 2D materials during the photolithographic process. This technique effectively prevents the 2D materials from being contaminated by photoresists, thereby ensuring pristine metal-semiconductor contacts. Through this innovative approach, we successfully fabricated monolayer $MoS_2$ transistors with pristine metal contacts, achieving high-quality electrical performance. We conducted large-scale fabrication using both traditional lithography and gold mask lithography. In the comparison of results, the average value for traditional lithography (TL) is 669 nA with a variance of $1.05 \times 10^6$ $nA^2$, whereas for gold mask lithography (GML), the average value is 33 nA with a variance of 250 $nA^2$. The application of this method resulted in devices with Ohmic contacts and an exceptionally low Schottky barrier height, both critical for optimal device operation.

The results highlight the stability of this metal mask technique introducing contamination-free 2D material transistors. This advancement not only enhances the performance and reliability of $MoS_2$-based devices but also opens new avenues for the development of integrated circuits utilizing 2D materials. The ability to fabricate high-quality, clean interfaces with minimal contamination is poised to make significant contributions to electronic device manufacturing and integration, paving the way for more sophisticated and efficient 2D material-based technologies.

Results and discussion

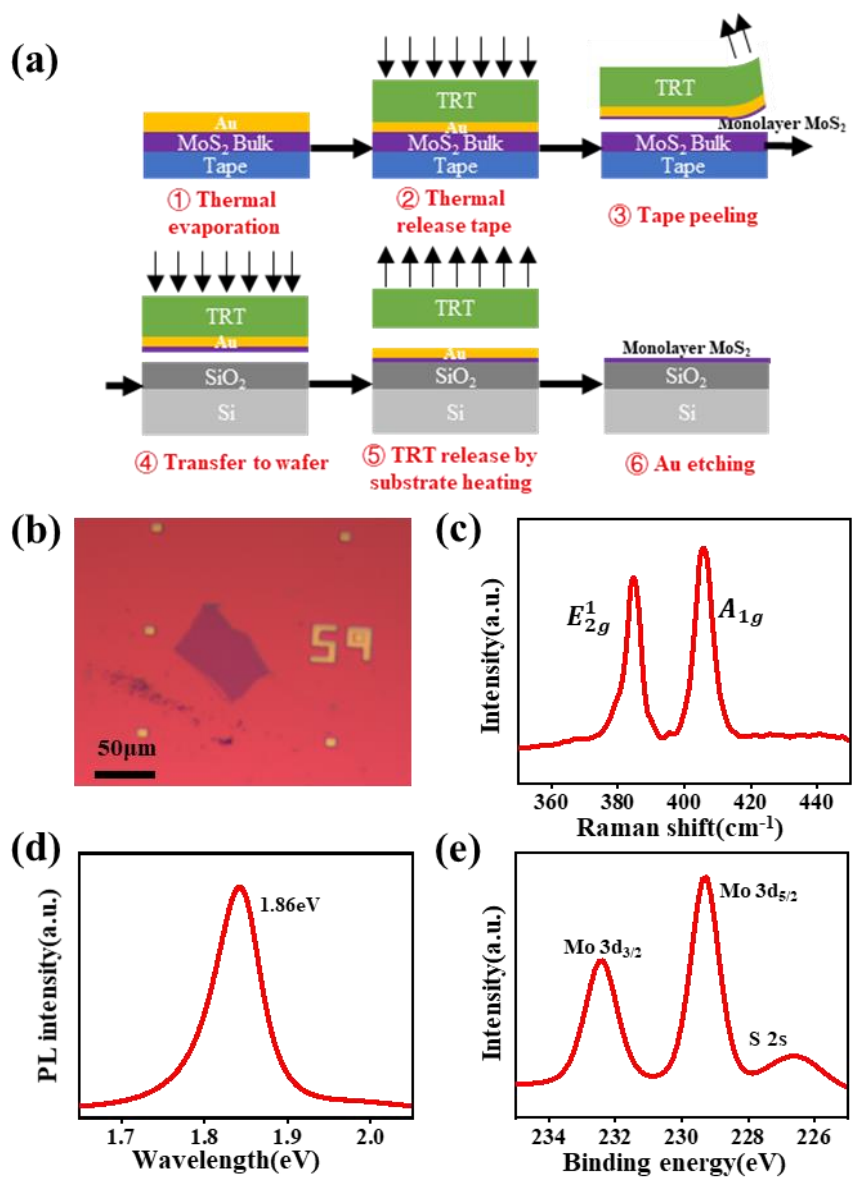

Figure 1. (a) Device process for the gold-assisted exfoliation to fabricate monolayer $MoS_2$. (b) Optical image of the monolayer $MoS_2$. (c) Photoluminescence spectrum of $MoS_2$ under green laser excitation (λ = 532 nm). (d) Raman spectroscopy of $MoS_2$. (e, f) XPS of Mo 3d and S 2p spectra for $MoS_2$.

A gold-assisted exfoliation technique was employed to exfoliate monolayers from bulk $MoS_2$[17], as illustrated in Fig.1(a). Initially, a thin layer of gold was thermally evaporated onto the fresh surface of a bulk crystal $MoS_2$ piece. Gold atoms will preferentially bond with the surface sulfur atoms. The Au-Si bonds are significantly stronger than the van der Waals force among $MoS_2$ layers, thereby facilitating the selective exfoliation of the top layer using thermal release tape. The tape was subsequently transferred to a target substrate ($SiO_2$/Si). The assembly was heated on a hotplate, followed by mild $O_2$ plasma treatment to remove any residual adhesive. The $O_2$ plasma power and etching duration were carefully controlled to prevent damage to the underlying monolayer $MoS_2$ during the etching process. Afterward, the gold film was selectively etched using a KI/I$_2$ solution, which effectively removes the gold without affecting the MoS$_2$.

The resultant MoS$_2$ monolayer are presented in Fig. 1(b). The monolayer nature of the MoS$_2$ was rigorously verified through advanced spectroscopic techniques. Raman spectroscopy provided definitive evidence of the monolayer status, indicated by a characteristic 18 cm$^{-1}$ peak separation between the E$_{2g}$ and A$_{1g}$ vibrational modes-a hallmark of monolayer MoS$_2$ (Fig.1c). This peak separation aligns with values reported in the literature, further confirming the successful isolation of a single MoS$_2$ layer[18].

Moreover, photoluminescence spectroscopy offered additional verification of the monolayer, revealing a prominent photoluminescence peak at 1.86 eV (Fig.1d) consistent with the direct bandgap of monolayer MoS$_2$[19]. This result underscores the high optical quality and uniformity of the exfoliated material. To complement these findings, X-ray photoelectron spectroscopy (XPS) was employed to analyze the chemical states and bonding environments of the MoS$_2$ (Fig.1e). These results collectively validate the successful fabrication of a high-quality and large-area MoS$_2$ monolayer[17].

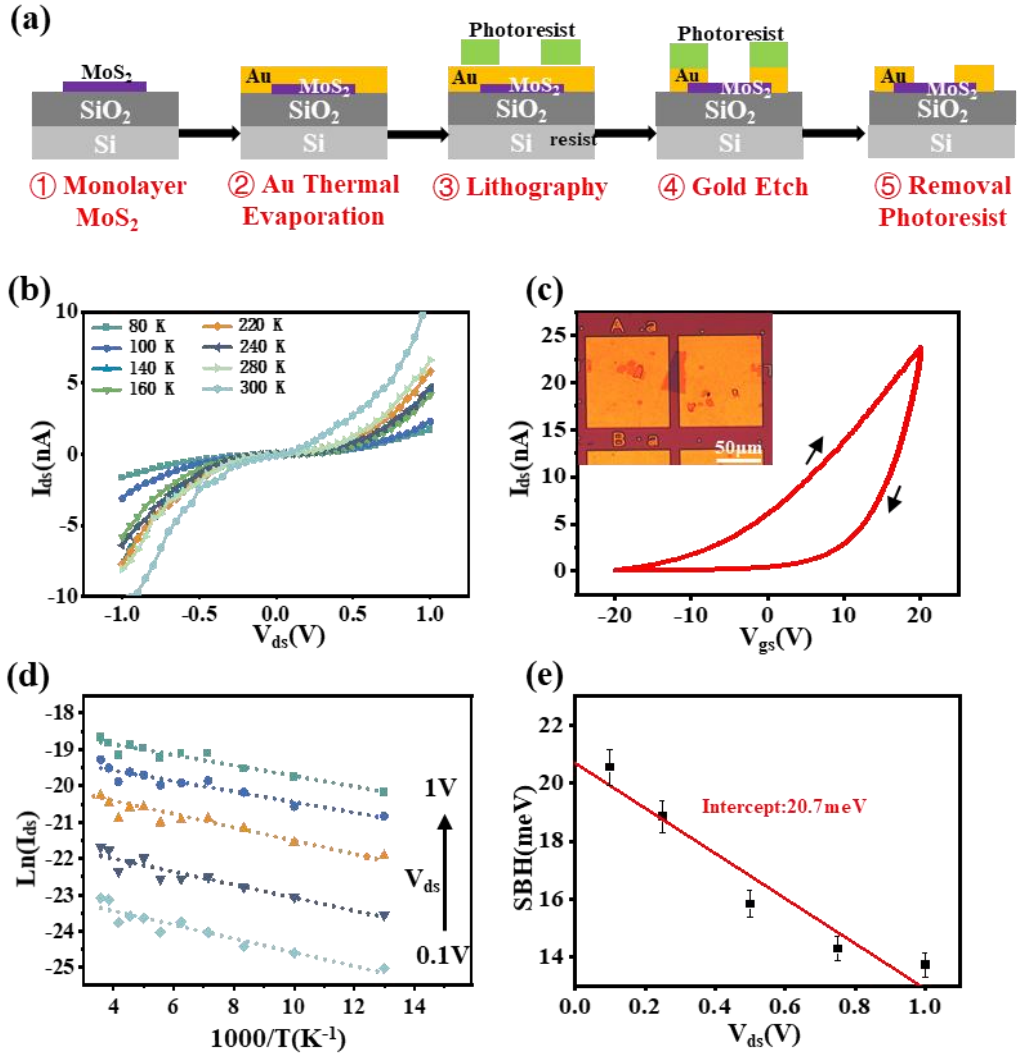

Figure 2. (a) Fabrication process for MoS$_2$ devices with pure gold electrode contact. (b) I-V curves at different temperatures. (c) Gate transfer characteristics of the MoS$_2$ transistor. Inset: Device optical microscopic image. (d) Linear fit of ln (I$_{ds}$) versus 1000/T. (e) Slopes extracted from the linear fit is plotted as a function of V$_{ds}$ and Φ$_B$ is derived from its y-intercept.

Fig. 2(a) depicts the fabrication process for the metal mask technique used with MoS$_2$. A 40 nm-thick layer of gold was thermally re-evaporated onto the exfoliated MoS$_2$, followed by the spin-coating of a photoresist. The photoresist made contact exclusively with the gold layer, avoiding direct interaction with the MoS$_2$. Standard photolithography was employed to pattern the photoresist. Gold etching was performed using a KI/I$_2$ solution, which selectively etched the gold without affecting the MoS$_2$ monolayer. The temperature-dependent I-V characteristics (80K to 300K), presented in Fig. 2(b), demonstrate an increase in current with rising temperature. The results indicate the presence of a strong Schottky barrier between the gold contact and MoS$_2$. The gate transfer characteristic curve at room temperature shown in Fig. 2(c) reveals the n-type transistor behavior of the MoS$_2$, with gate hysteresis attributed to defects and other non-ideal factors. The inset is the optical microscopic image of the device. Additionally, for the semiconductors that has a low carrier mobility like monolayer MoS$_2$, the Schottky junction is best described by the diffusion model. In this model, the saturation current at reverse bias is proportional to electric field intensity and the exponential term of Schottky barrier height $e^{-\frac{q}{k_B T}\Phi_B}$.

$$I_{ds} = AA_{2D}^{**} \exp\left[-\frac{q}{k_B T}\left(\Phi_B - \frac{V_{ds}}{n}\right)\right] \tag{1}$$

, where A represents the contact area, $A_{2D}^{**}$ represents the effective two-dimensional equivalent Richardson constant which is temperature independent, q denotes the magnitude of the electron charge, k$_B$ is the Boltzmann constant, Φ$_B$ refers to the Schottky barrier height, n is the ideality factor, and V$_{ds}$ is the source-drain bias. As shown in Fig. 2(d), ln (I$_{ds}$) is plotted against 1000/T under different source-drain bias (V$_{ds}$). The data is linear at each bias, and the Schottky barrier height was calculated from the slope equal to $-\frac{q}{k_B T}\Phi_B$, subsequently the Φ$_B$ was plotted in Fig. 2(e) as a function of drain-source bias. The intercept of the fitted linear curves of the plot in Fig. 2(e) means the Φ$_B$ is 20.7meV. As shown in Fig. 2(e), the Schottky barrier height decreases with increasing bias voltage. This reduction can be attributed to the modulation effect of the Schottky barrier, where higher bias allows carriers to more easily overcome the barrier.

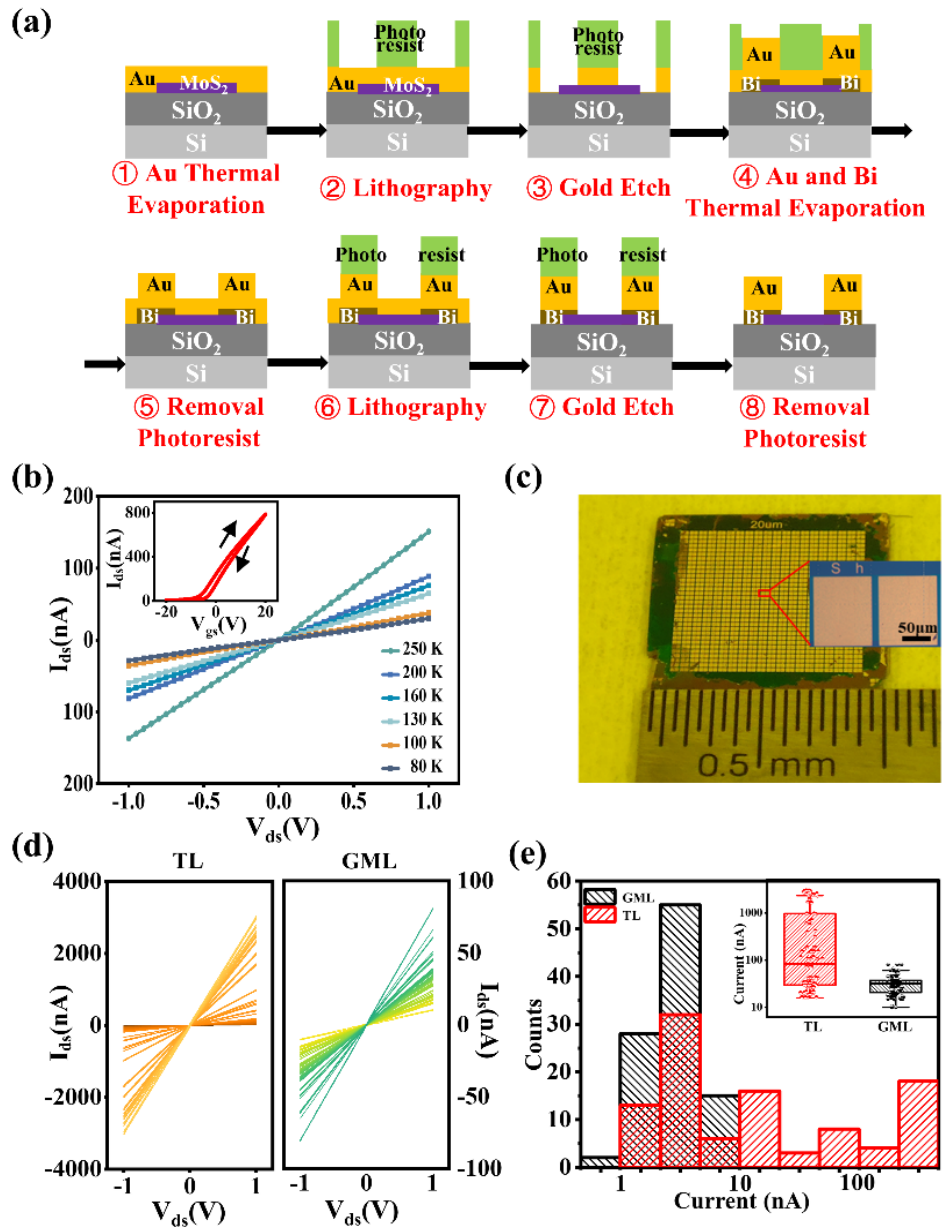

Figure 3. (a) Fabrication process for MoS$_2$ devices with Bismuth and gold electrode contact. (b) I-V curves in different temperatures. Inset: gate transfer characteristics of the MoS$_2$ transistor. (c) Optical microscope image of device array fabricated by Gold Mask Lithography, with the magnified view of a device. (d) I-V curves of traditional lithography and gold mask lithography for 100 devices. (e) Current Statistics of the device fabricated by classical photolithography and our gold masked lithography.

However, metals other than gold are often required to make Ohmic contacts with monolayers MoS$_2$[20], [21], [22]. For example, Bismuth, known for its low melting point and semi-metallic characteristics, is a suitable metal to form Ohmic contacts with the MoS$_2$ monolayer[23]. To facilitate this possibility, we propose a new paradigm on top of the previous process in which any metals can make a clean contact with the 2D monolayer without the contamination of the polymeric photoresist. As shown in Fig. 3(a), gold is first thermally evaporated onto the 2D material surface, followed by the spin-coating of photoresist. Photolithography and metal etching are subsequently performed. The pattern in the photoresist is transferred to the protecting gold film. The desired metal films (Bi and Au in our case) are then evaporated to contact the 2D materials. The remaining metal films are removed in the following lift-off process. Photolithography with a pattern complementary to the previous pattern is performed to protect the desired metal films from subsequent gold etching. Throughout the entire processes, $MoS_2$ remained clean without direct contact with photoresist.

Fig.3(b) presents the current-voltage (I-V) characteristics of the device measured at different temperatures. In comparison with the I-V characteristics shown in Fig.2(b), the device exhibits a linear I-V relationship from room temperature down to 80K, indicating the formation of a high-quality Ohmic contact. The consistent linearity of the I-V curves under different thermal conditions further demonstrates the stability and reliability of the contact interface, highlighting the effectiveness of the fabrication process in achieving optimal electrical performance. The inset is the gate transfer characteristics of the $MoS_2$ transistor, showing that the $MoS_2$ channel is n-type. The narrow hysteretic loop indicates that the defects in these $MoS_2$ devices are significantly suppressed.

To compare the classical photolithography and the gold mask lithography, we fabricated 100 Au/Bi contact MoS2 devices with each of these two lithography technologies with 200 devices in total. To ensure the consistency in the experimental conditions, all of the devices were fabricated with identical geometries and electrode thicknesses. Fig. 3(c) shows the optical image of the $MoS_2$ device array on the chip. The inset shows the optical microscopic image of a single device.

Fig. 3(d) presents I-V curves of traditional lithography and gold mask lithography for 100 devices separately. The currents in darkness at a constant bias of 1V applied across the devices are shown in Fig. 3(e), The small figure is a box plot of the data, displaying the interquartile range (IQR) from the 25th to the 75th percentiles, with the central line representing the median. The whiskers extend to the minimum and maximum values within 1.5 times the IQR. The current of the devices produced by classical photolithography varies from device to device in a range of 2-orders-of-magnitude (from tens of nA to several μA). This variability is attributed to the uncontrollable organic doping effects of the photoresist residue to the $MoS_2$ material. In contrast, The current of the devices produced by the gold mask photolithography concentrate within one order of magnitude at ~ tens of nA. Clearly, the gold mask lithography effectively mitigates the doping effect of the photoresist residue, thereby enhancing the stability and reliability of device performances. These devices are fabricated in the cleanroom at a class of ~100,000. We believe that the variation from device to device fabricated by the gold mask lithography will be much smaller if these devices are fabricated in a lower-class cleanroom.

Conclusion

In this study, we successfully developed a novel fabrication process for monolayer MoS$_2$ transistors, utilizing gold as a protective mask during the patterning and etching stages. This technique effectively mitigates contamination from photoresists and other organic solvents, preserving the pristine condition of the MoS$_2$ surface and ensuring high-quality metal contacts. As a result, the fabricated transistors exhibit Ohmic contacts with minimal Schottky barrier height and robust electrical performance. This advancement not only significantly enhances the performance and reliability of MoS$_2$-based transistors but also establishes a benchmark for contamination-free fabrication of 2D material devices. Consequently, it paves the way for their integration into high-performance electronic and optoelectronic applications, potentially advancing technology in these fields.

Reference

[1] "Effect of substrate and temperature on the electronic properties of monolayer molybdenum disulfide field-effect transistors," *Physics Letters A*, vol. 382, no. 10, pp. 697–703, Mar. 2018, doi: 10.1016/j.physleta.2017.12.052.

[2] Y. Liu and F. Gu, "A wafer-scale synthesis of monolayer MoS2 and their field-effect transistors toward practical applications," *Nanoscale Adv.*, vol. 3, no. 8, pp. 2117–2138, Apr. 2021, doi: 10.1039/D0NA01043J.

[3] "Molybdenum Disulfide-based field effect transistor biosensors for medical diagnostics: Exploring a decade of advancements (2014–2024)," *TrAC Trends in Analytical Chemistry*, vol. 176, p. 117742, Jul. 2024, doi: 10.1016/j.trac.2024.117742.

[4] L.-R. Zou *et al.*, "Research progress of optoelectronic devices based on two-dimensional MoS2 materials," *Rare Met.*, vol. 42, no. 1, pp. 17–38, Jan. 2023, doi: 10.1007/s12598-022-02113-y.

[5] A. Nourbakhsh *et al.*, "MoS$_2$ Field-Effect Transistor with Sub-10 nm Channel Length," *Nano Lett.*, vol. 16, no. 12, pp. 7798–7806, Dec. 2016, doi: 10.1021/acs.nanolett.6b03999.

[6] A. Khare and P. Dwivedi, "Design, simulation and optimization of multi-layered MoS$_2$ based FET devices," *Eng. Res. Express*, Nov. 2021, doi: 10.1088/2631-8695/ac3d11.

[7] "A non-destructive method for the removal of residual resist in imprinted patterns," *Microelectronic Engineering*, vol. 67–68, pp. 245–251, Jun. 2003, doi: 10.1016/S0167-9317(03)00184-9.

[8] W. Den, S.-C. Hu, C. M. Garza, and O. A. Zargar, "Review—Airborne Molecular Contamination: Recent Developments in the Understanding and Minimization for Advanced Semiconductor Device Manufacturing," *ECS J. Solid State Sci. Technol.*, vol. 9, no. 6, p. 064003, Jul. 2020, doi: 10.1149/2162-8777/aba080.

[9] H. J. Jang, J. Y. Kim, E. Y. Jung, M. Choi, and H.-S. Tae, "Photoresist Removal Using Reactive Oxygen Species Produced by an Atmospheric Pressure Plasma Reactor," *ECS J. Solid State Sci. Technol.*, vol. 11, no. 4, p. 045010, Apr. 2022, doi: 10.1149/2162-8777/ac62ef.

[10] A. West, M. van der Schans, C. Xu, M. Cooke, and E. Wagenaars, "Fast, downstream removal of photoresist using reactive oxygen species from the effluent of an atmospheric pressure plasma Jet," *Plasma Sources Sci. Technol.*, vol. 25, no. 2, p. 02LT01, Mar. 2016, doi:

10.1088/0963-0252/25/2/02LT01.

[11] M. T. Pettes, I. Jo, Z. Yao, and L. Shi, "Influence of Polymeric Residue on the Thermal Conductivity of Suspended Bilayer Graphene," *Nano Lett.*, vol. 11, no. 3, pp. 1195–1200, Mar. 2011, doi: 10.1021/nl104156y.

[12] Y. Dan, Y. Lu, N. J. Kybert, Z. Luo, and A. T. C. Johnson, "Intrinsic Response of Graphene Vapor Sensors," *Nano Lett.*, vol. 9, no. 4, pp. 1472–1475, Apr. 2009, doi: 10.1021/nl8033637.

[13] "Improving consistency and performance of graphene-based devices via Al sacrificial layer," *Colloid and Interface Science Communications*, vol. 56, p. 100743, Sep. 2023, doi: 10.1016/j.colcom.2023.100743.

[14] Y.-C. Lin, C.-C. Lu, C.-H. Yeh, C. Jin, K. Suenaga, and P.-W. Chiu, "Graphene Annealing: How Clean Can It Be?," *Nano Lett.*, vol. 12, no. 1, pp. 414–419, Jan. 2012, doi: 10.1021/nl203733r.

[15] T. Kwon, H. An, Y.-S. Seo, and J. Jung, "Plasma Treatment to Improve Chemical Vapor Deposition-Grown Graphene to Metal Electrode Contact," *Jpn. J. Appl. Phys.*, vol. 51, no. 4S, p. 04DN04, Apr. 2012, doi: 10.1143/JJAP.51.04DN04.

[16] "Efficient cleaning of graphene from residual lithographic polymers by ozone treatment," *Carbon*, vol. 109, pp. 221–226, Nov. 2016, doi: 10.1016/j.carbon.2016.08.013.

[17] S. B. Desai *et al.*, "Gold-Mediated Exfoliation of Ultralarge Optoelectronically-Perfect Monolayers," *Advanced Materials*, vol. 28, no. 21, pp. 4053–4058, 2016, doi: 10.1002/adma.201506171.

[18] S. Zhang *et al.*, "Direct Observation of Degenerate Two-Photon Absorption and Its Saturation in WS2 and MoS2 Monolayer and Few-Layer Films," *ACS Nano*, vol. 9, no. 7, pp. 7142–7150, Jul. 2015, doi: 10.1021/acsnano.5b03480.

[19] K. F. Mak, C. Lee, J. Hone, J. Shan, and T. F. Heinz, "Atomically Thin ${\mathrm{MoS}}_{2}$: A New Direct-Gap Semiconductor," *Phys. Rev. Lett.*, vol. 105, no. 13, p. 136805, Sep. 2010, doi: 10.1103/PhysRevLett.105.136805.

[20] W. Li *et al.*, "Approaching the quantum limit in two-dimensional semiconductor contacts," *Nature*, vol. 613, no. 7943, pp. 274–279, Jan. 2023, doi: 10.1038/s41586-022-05431-4.

[21] S. Lee *et al.*, "Semi-Metal Edge Contact for Barrier-Free Carrier Transport in MoS2 Field Effect Transistors," *ACS Appl. Electron. Mater.*, vol. 6, no. 6, pp. 4149–4158, Jun. 2024, doi: 10.1021/acsaelm.4c00250.

[22] Z. Liu, Q. Zhang, X. Huang, C. Liu, and P. Zhou, "Contact engineering for temperature stability improvement of Bi-contacted MoS2 field effect transistors," *Sci. China Inf. Sci.*, vol. 67, no. 6, p. 160402, May 2024, doi: 10.1007/s11432-023-3942-2.

[23] P.-C. Shen *et al.*, "Ultralow contact resistance between semimetal and monolayer semiconductors," *Nature*, vol. 593, no. 7858, pp. 211–217, May 2021, doi: 10.1038/s41586-021-03472-9.